\documentclass[pra,aps,twocolumn,superscriptaddress,nofootinbib,nopacs]{revtex4} 

\usepackage{amsmath}  \usepackage{amssymb}  \usepackage{amsfonts}  \usepackage{bm}  \usepackage{bbm}   \usepackage{braket}  \usepackage{color}  \usepackage{comment}  \usepackage{dcolumn}  \usepackage{enumerate}  \usepackage{epsfig}  \usepackage{gensymb}  \usepackage{graphicx}  \usepackage{indentfirst}  \usepackage{lmodern}  \usepackage{mathrsfs}  \usepackage{mathtools}  \usepackage{psfrag}  \usepackage{pst-all}  \usepackage{soul}  \usepackage{units}  \usepackage{xcolor}
\usepackage{float} 
\usepackage[colorlinks,linkcolor=blue,citecolor=blue,urlcolor=blue,hyperindex,driverfallback=dvipdfm]{hyperref}  \usepackage[T1]{fontenc} 

\def\ii{{\rm i}}  \def\ee{{\rm e}}
\def\me{m_{\rm e}}  
  
    \def\Bb{{\bf B}}    \def\Eb{{\bf E}}                  \def\jb{{\bf j}}                  \def\Rb{{\bf R}}  \def\rb{{\bf r}}       
  \def\yy{\hat{\bf y}}  \def\zz{\hat{\bf z}}            
\def\kpar{k_\parallel}  \def\kparb{{\bf k}_\parallel} 
      
\begin{document} 

\title{Inelastic Scattering of Electron Beams by Nonreciprocal Nanotructures}


\author{Renwen~Yu}
\affiliation{ICFO-Institut de Ciencies Fotoniques, The Barcelona Institute of Science and Technology, 08860 Castelldefels (Barcelona), Spain}
\author{Andrea~Kone\v{c}n\'{a}}
\affiliation{ICFO-Institut de Ciencies Fotoniques, The Barcelona Institute of Science and Technology, 08860 Castelldefels (Barcelona), Spain}
\author{F.~Javier~Garc\'{\i}a~de~Abajo}
\email{javier.garciadeabajo@nanophotonics.es}
\affiliation{ICFO-Institut de Ciencies Fotoniques, The Barcelona Institute of Science and Technology, 08860 Castelldefels (Barcelona), Spain}
\affiliation{ICREA-Instituci\'o Catalana de Recerca i Estudis Avan\c{c}ats, Passeig Llu\'{\i}s Companys 23, 08010 Barcelona, Spain}


\begin{abstract}
Probing optical excitations with high resolution is important for understanding their dynamics and controlling their interaction with other photonic elements. This can be done using state-of-the-art electron microscopes, which provide the means to sample optical excitations with combined meV--sub-nm energy--space resolution. For reciprocal photonic systems, electrons traveling in opposite directions produce identical signals, while this symmetry is broken in nonreciprocal structures. Here, we theoretically investigate this phenomenon by analyzing electron energy-loss spectroscopy (EELS) and cathodoluminescence (CL) in structures consisting of magnetically biased InAs as an instance of gyrotropic nonreciprocal material. We find that the spectral features associated with excitations of InAs films depend on the electron propagation direction in both EELS and CL, and can be tuned by varying the applied magnetic field within a relatively modest sub-tesla regime. The magnetic field modifies the optical field distribution of the sampled resonances, and this in turn produces a direction-dependent coupling to the electron. The present results pave the way to the use of electron microscope spectroscopies to explore the near-field characteristics of nonreciprocal systems with high spatial resolution.
\end{abstract}

\maketitle 
\date{\today} 

\section*{INTRODUCTION}

Electron energy-loss spectroscopy (EELS) performed in scanning transmission electron microscopes is a powerful technique to investigate the spatial and spectral characteristics of materials excitations over a wide range of energies \cite{E96,E03,H03,EB05,B06,MBM07,paper149,KS14,paper338,paper371}. More precisely, detailed information on the chemical composition of material structures is routinely gathered by monitoring high-energy losses using this technique \cite{MTR93,E96,MKM08,BDK02}, while the low-loss region of the EELS spectra provides an unsurpassed way of spatially mapping plasmons \cite{BKW07,paper085,RB13,KS14,paper369}, phonons and phonon polaritons \cite{KLD14,SSB19,HRK20,paper361,YLG21}, and excitons \cite{BLS21}. State-of-the-art instruments are currently capable of delivering a combined spectral and spatial resolution in the few-meV and sub-nm range, which allows mapping mid-infrared excitations \cite{KLD14,LTH17,LB18,HNY18,HHP19,HKR19,paper342,HRK20,YLG21,paper369}. In addition, the cathodoluminescence (CL) emission associated with electron-driven excitations of optically active modes can equally provide spatially resolved imaging without the requirement of having electron-transparent samples \cite{paper035,BJK06,VDK07,paper167,MCW18,paper341}. In general, the intensities collected in both EELS and CL depend on the electron trajectory relative to the specimen, and in particular, in virtue of the reciprocity theorem, the EELS probability remains unchanged when reversing the direction of the electron velocity if the sampled structure is made of reciprocal media, while CL is also invariant for systems that possess inversion symmetry along the electron beam (e-beam) direction.

Nonreciprocal photonic structures are currently attracting much attention because they provide appealing ways to control the propagation of electromagnetic waves \cite{WF05,YVW08,HR08,LJS14}. In particular, gyrotropic materials enable the development of optical circulators based on the Cotton-Mouton effect \cite{ST19}, as well as the realization of one-way electromagnetic wave flow \cite{YVW08,LJS14} and the manifestation of exotic phenomena such as the violation of detailed balance in the emission of thermal radiation \cite{ZF14}. When exposed to an external DC magnetic field, gyrotropic media display a nonreciprocal response that emanates from the cyclotron orbits described by electrons in the bulk of the material, which result in an anisotropic permittivity tensor with complex off-diagonal components. In a related context, the nonreciprocity arising from a graphene layer that hosts a drift electrical current has been predicted to produce a dependence on the sign of the e-beam direction in Cherenkov \cite{PS18} and EELS \cite{PS21} spectra for free electrons passing near the material with relatively low speeds of the order of the Fermi velocity \cite{PS18}. Additionally, the presence of an intense external e-beam can trigger nonreciprocal behavior in the guided modes of a metallic cavity \cite{FKS21}. These studies capitalize on the ability of fast electrons to interact with optical modes that do not couple to far-field radiation, but still, a trace of their excitation is directly revealed in the electron-generated spectra. We expect that e-beams can be used to probe the nonreciprocal response with high spatial resolution through the mechanism of near-field coupling to the sample, although this possibility remains so far unexplored.

Here, we predict a large dependence of the EELS and CL spectra on both the strength of an externally applied DC magnetic field and the sign of the probe velocity vector for free electrons interacting with nonreciprocal waveguides that incorporate gyrotropic materials. More precisely, we explore this phenomenon by examining the terahertz spectra associated with planar InAs films, the waveguided modes of which manifest as sharp spectral EELS features undergoing sizeable shifts when applying tesla-scale magnetic fields, such as those provided by permanent magnets \cite{F01}. Similar effects are observed in the CL emission resulting from out-coupling of the waveguide modes through a surface grating. Electrons are thus sensitive to the strength of the magnetic field when probing high-quality-factor resonances in nonreciprocal structures made of gyrotropic materials.

\section*{RESULTS AND DISCUSSION}

Taking the electron to move with constant velocity $v$ along $z$, following a straight line trajectory determined by $\Rb_0=(x_0,y_0)$, the spectrally resolved EELS probability is given by \cite{paper149,paper371}
\begin{align}
&\Gamma_{{\rm EELS}}(\omega) \nonumber\\ 
&=-\frac{4e^{2}}{\hbar}\int dz\int dz'{\rm Im}\left\{ {\rm e}^{{\rm i}\omega(z'-z)/v} G_{zz}\left(\mathbf{R}_{0},z,\mathbf{R}_{0},z',\omega\right)\right\} \nonumber
\end{align}
in terms of the $zz$ component of the electromagnetic Green tensor $G(\rb,\rb',\omega)$, implicitly defined through the relation $\nabla\times\nabla\times G(\rb,\rb',\omega)-k^2\epsilon(\rb,\omega)\cdot G(\rb,\rb',\omega)=(-1/c^2)\delta(\rb-\rb')$, where $\epsilon(\rb,\omega)$ is the position- and frequency-dependent local permittivity tensor, while $k=\omega/c$ is the free-space light wave vector. In structures made of reciprocal materials, the Green tensor satisfies the property $G_{zz}(\rb,\rb',\omega)=G_{zz}(\rb',\rb,\omega)$, which directly leads to the invariance of $\Gamma_{{\rm EELS}}(\omega)$ under the transformation $v\rightarrow-v$ (i.e., reversing the direction of the electron velocity vector). However, in the presence of nonreciprocal materials, the loss probability depends on the sign of $v$.

We consider structures containing indium arsenide (InAs) as an example of material that exhibits a pronounced nonreciprocal response when exposed to a magnetic DC field $\Bb$ \cite{ZF14}. Using Cartesian coordinates and taking the magnetic field oriented as $\Bb=B\yy$, the permittivity tensor of InAs becomes \cite{ZF14}
\begin{align}
\epsilon_{\rm InAs}(\omega)=\left[\begin{array}{ccc}
\epsilon_{xx}(\omega) & 0 & \epsilon_{xz}(\omega) \\
0 & \epsilon_{yy}(\omega) & 0 \\
\epsilon_{zx}(\omega) & 0 & \epsilon_{zz}(\omega)
\end{array}\right],
\nonumber
\end{align}
where
\begin{align}
\epsilon_{xx}(\omega)&=\epsilon_{zz}(\omega)=\epsilon_{\infty}-\frac{\omega_{{\rm p}}^{2}(\omega+{\rm i}\gamma)}{\omega\left[(\omega+{\rm i}\gamma)^{2}-\omega_{{\rm c}}^{2}\right]},  \nonumber\\
\epsilon_{xz}(\omega)&=-\epsilon_{zx}(\omega)=\frac{\ii\omega_{{\rm p}}^{2}\omega_{{\rm c}}}{\omega\left[(\omega+{\rm i}\gamma)^{2}-\omega_{{\rm c}}^{2}\right]}, \nonumber\\
\epsilon_{yy}(\omega)&=\epsilon_{\infty}-\frac{\omega_{{\rm p}}^{2}}{\omega(\omega+{\rm i}\gamma)},
\nonumber
\end{align}
with $\epsilon_{\infty}=12.37$, $\hbar\omega_{{\rm p}}=180.53\,$meV, and $\hbar\gamma=102.02\,$\textmu eV. The magnetic field enters through the cyclotron frequency $\omega_{{\rm c}}=eB/m^{*}$, where $m^{*}=0.33\,\me$ is the effective electron mass in this material. Our structures also contain aluminum (Al) or silicon carbide (SiC), which we describe through their isotropic, frequency-dependent, local permittivities, taken from tabulated optical measurements \cite{P1985}.

For planar structures, we obtain the loss probability by expressing the Green tensor in terms of the Fresnel reflection coefficients $r_{\sigma\sigma'}$, where $\sigma$ and $\sigma'$ run over p and s polarization components. Cross-polarization terms proportional to $r_{\rm sp}$ and $r_{\rm ps}$ emerge as a result of the off-diagonal elements of $\epsilon_{\rm InAs}(\omega)$ produced in the presence of a magnetic field. Taking the surface at the $x=0$ plane, we find \cite{paper045}
\begin{align}
&\Gamma_{{\rm EELS}}(\omega)=\frac{e^2L}{\pi\hbar v^2}\int \frac{dk_y}{\kpar^2}\,\ee^{-2\kappa x_0} \label{EELSr}\\
&\times\left(\kappa\,{\rm Im}\left\{
r_{\rm pp}\right\}+\frac{k_y^2v^2}{\kappa c^2}\,{\rm Im}\left\{r_{\rm ss}\right\}+\frac{k_yv}{c}{\rm Re}\left\{r_{\rm sp}-r_{\rm ps}
\right\}\right) \nonumber,
\end{align}
where $L$ is the length of the electron trajectory, $x_0$ is the electron-surface separation, $\kappa=\sqrt{(\omega/v\gamma)^2+k_y^2}$ with $\gamma=1/\sqrt{1-v^2/c^2}$, the Fresnel coefficients depend on the loss frequency $\omega$ and the parallel wave vector transfer $\kparb=(k_y,k_z)$, and momentum conservation along the beam imposes the condition $k_z=\omega/v$. For lossless dielectric waveguides, the integrand in Eq.\ (\ref{EELSr}) vanishes, except at the guided modes, which are signalled by singularities of the Fresnel coefficients. Analytical results for the Fresnel coefficients become too intricate, so we resort instead on numerical simulations in what follows, with the EELS probability $\Gamma_{\rm EELS}(\omega)=(e/\pi\hbar\omega)\int d^3\rb\;{\rm Re}\{\jb^{\rm ext *}(\rb,\omega)\cdot\Eb^{\rm ind}(\rb,\omega)\}$ expressed in terms of the self-induced electric field \cite{paper149} $\Eb^{\rm ind}(\rb,\omega)$, which is in turn calculated using a frequency-domain finite-difference electromagnetic solver (COMSOL) with the electron source introduced through a line current $\jb^{\rm ext}(\rb,\omega)=-e\zz\,\delta(x-x_0)\delta(y)\ee^{\ii\omega z/v}$. The CL probability is also obtained numerically from the far-field amplitude for the same external source.

\begin{figure}
\centering{\includegraphics[width=0.5\textwidth]{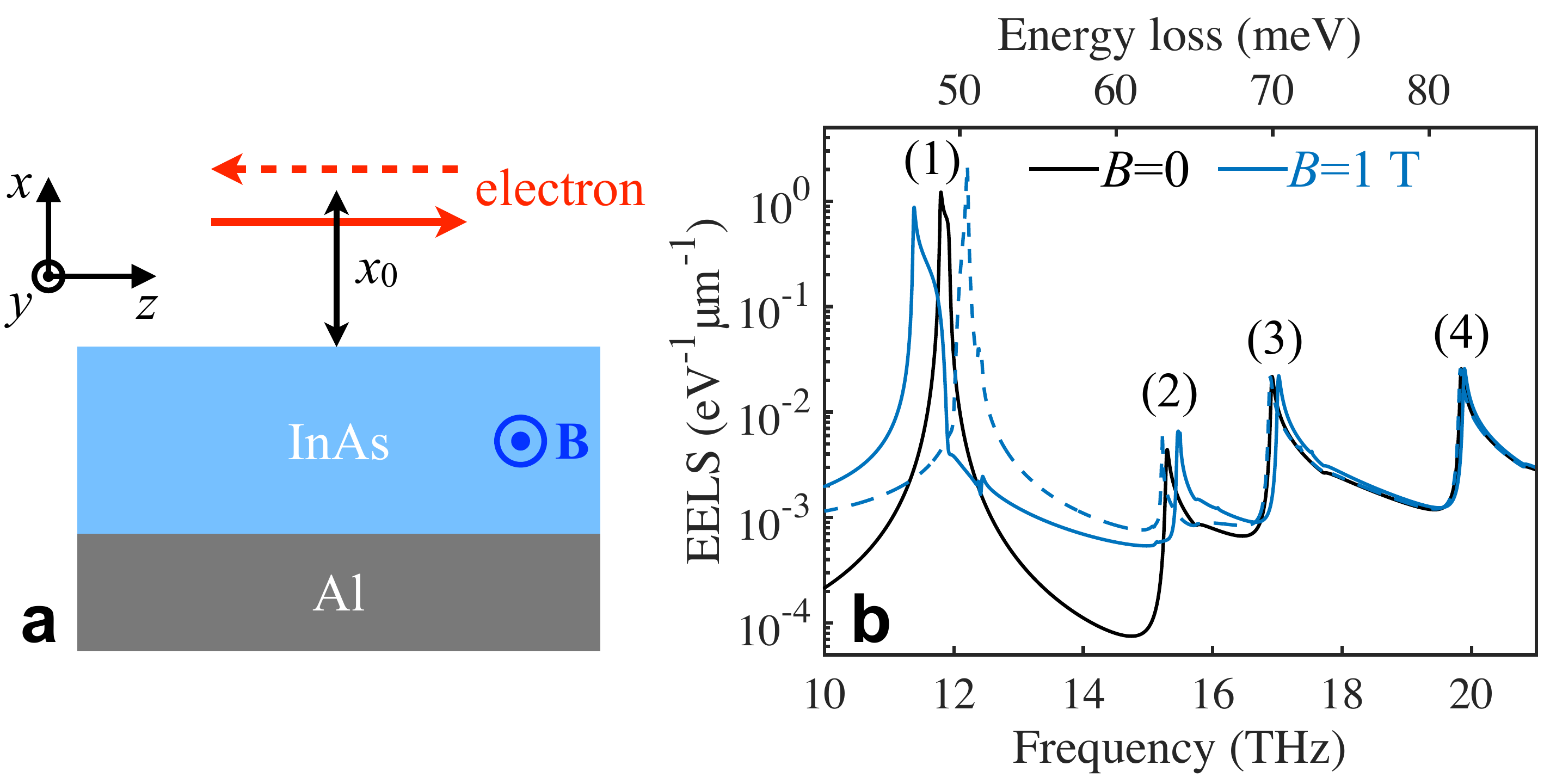}}
\caption{{\bf Aloof electron-beam interaction with a nonreciprocal semiconductor film.} {\bf (a)} We consider electrons moving parallel to an InAs planar film supported on Al and exposed to an in-plane magnetic field $\Bb$ with the orientation indicated in the figure. The electron-surface interaction depends on the direction of the electron velocity due to the nonreciprocal response induced by the magnetic field in InAs. {\bf (b)} Simulated EELS probability spectra for electrons moving to the left (dashed curves) or to the right (solid curves) with (blue, $B=1\,$T) and without (black, $B=0$) an applied magnetic field under the geometry sketched in (a). The electron-surface separation is 1\,{\textmu}m, the InAs layer thickness is 9.85\,{\textmu}m, and the electron velocity is $v=0.5\,c$. Features (1)-(4) label different guided modes excited by the electron.}
\label{Fig1}
\end{figure}

Without external magnetic fields, we find four resonance peaks in the EELS spectra simulated from Eq.\ (\ref{EELSr}) [black curve in Fig.\ \ref{Fig1}(b)]. The first peak around 12\,THz (50\,meV in energy scale) is associated with the surface plasmon resonances supported at the air-InAs interface, whereas the remaining peaks represent the excitation of the bulk waveguide modes contained in the planar InAs film. In the presence of an external magnetic field of strength $B=1\,$T, each of these resonance peaks shifts to the red and blue for electrons traveling along positive (solid blue curve) and negative (dashed blue curve) \textit{z}-directions, respectively, as shown in Fig.\ \ref{Fig1}(b), thus revealing a nonreciprocal response with respect to the electron propagation direction. Incidentally, the electron can be deflected by the magnetic field, but such deflection does not affect the electron spectrum and its magnitude is negligible for electron-sample interaction lengths in the range of a few microns.

\begin{figure}
\centering{\includegraphics[width=0.5\textwidth]{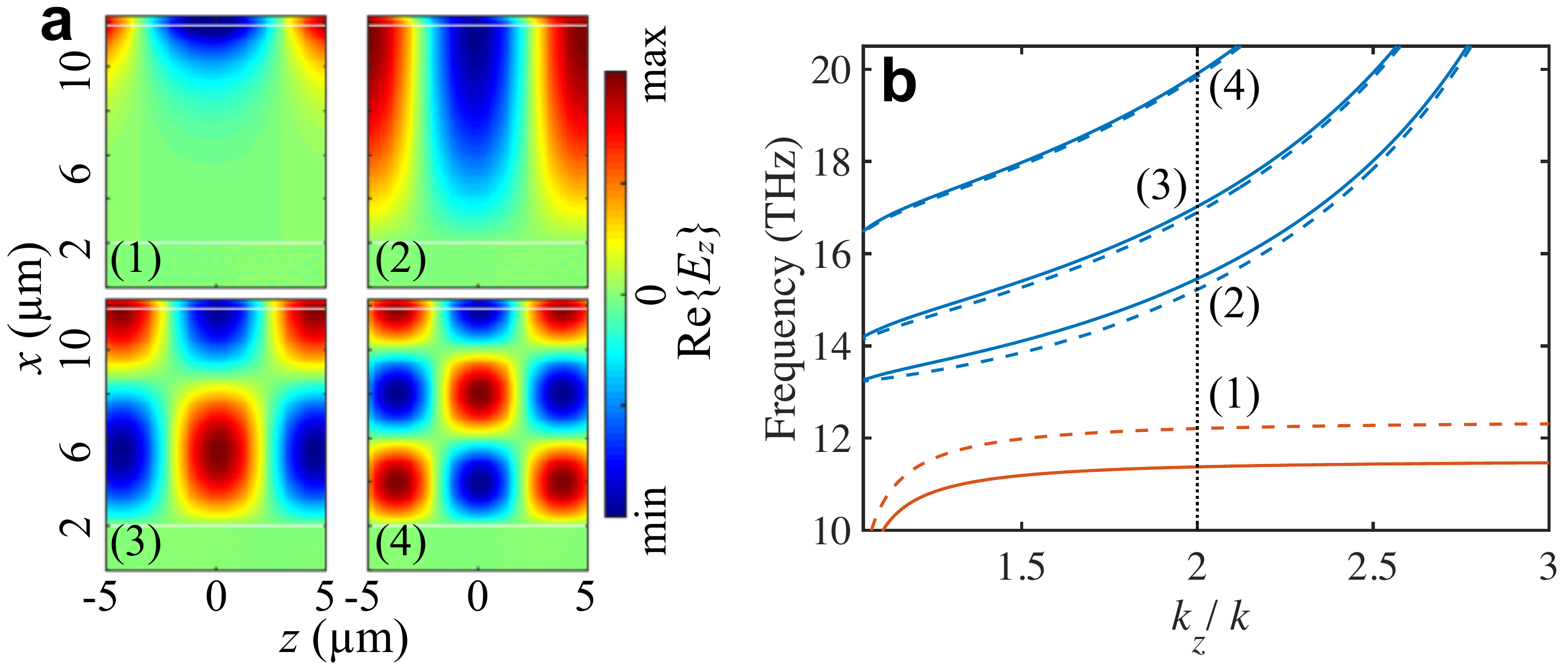}}
\caption{{\bf Guided modes in a nonreciprocal semiconductor film.} {\bf (a)} Electric near-field amplitude distributions corresponding to the resonances (1)-(4) shown in Fig.\ \ref{Fig1}(b) for $B=0$. The upper and lower solid white lines indicate the air-InAs and InAs-Al interfaces, respectively. {\bf (b)} Dispersion diagrams of guided surface (red) and bulk (blue) modes supported by the film shown in Fig.\ \ref{Fig1}(a) with $B=1\,$T. The in-plane wave vector $k_z$ is normalized to the light wave vector $k=\omega/c$. Dashed curves are the mirror reflection of the dispersion diagrams for $k_z<0$, which are found to be shifted with respect to those for $k_z>0$ (solid curves). We show the electron line $\omega=k_zv$ with $v=0.5\,c$ for reference (vertical dotted line).}
\label{Fig2}
\end{figure}

To explore the origin of the four resonance peaks observed in Fig.\ \ref{Fig1}(b), we plot the electric field distributions of their associated modes in Fig.\ \ref{Fig2}(a). For the first resonance peak (label (1) in both Figs.\ \ref{Fig1}(b) and \ref{Fig2}(a) at $B=0$), the field is confined at the upper air-InAs interface (indicated by the upper horizontal white line), demonstrating its surface-mode nature. In contrast, the electric fields of the three remaining resonance modes [labels (2)-(4) in Fig.\ \ref{Fig2}(a) at $B=0$], penetrate significantly inside the InAs film (region in between the two horizontal white lines) and form standing wave patterns along the \textit{z} direction, a behavior that is typical of TM-polarized waveguide modes (these are in fact the $\rm TM_0$, $\rm TM_1$, and $\rm TM_2$ modes of the planar dielectric film \cite{SL12}). In the presence of a magnetic field, the optical near-field is distorted and lacks inversion symmetry relative to the $z=0$ plane when the sign of $k_z$ is reversed; the changes in the optical field are also confined near the surface in mode (1) and extended across the bulk of the film in modes (2)-(4), as shown in supplementary Fig.\ \ref{FigS1}. In Fig.\ \ref{Fig2}(b), we further plot the asymmetric dispersion relations of the guided modes supported in the structure displayed in Fig.\ \ref{Fig1}(a) for positive (solid curves) and negative (dashed curves, mirror-reflected with respect to $k_z=0$) values of $k_z$ at $k_y=0$. For simplicity, the results in Fig.\ \ref{Fig2}(b) are obtained by assuming the Al substrate to respond as a perfect electric conductor, although the rest of our calculations take into consideration the metal dielectric function, as indicated above. The waveguide dispersion relations for $k_y=0$ are then given by the expression \cite{ZF14}
\begin{align}
\tan\left(\alpha\right)=\frac{\epsilon_{zz}\sqrt{\epsilon_{1}-\beta^{2}}}{\left(\epsilon_{zz}-\beta^{2}\right)/\sqrt{\beta^{2}-1}+\ii\epsilon_{xz}\beta}, \nonumber
\end{align}
where $\epsilon_{1}=\left(\epsilon_{zz}^{2}+\epsilon_{xz}^{2}\right)/\epsilon_{zz}$,
$\beta=k_{z}c/\omega$, and $\alpha=\left(\omega\sqrt{\epsilon_{1}-\beta^{2}}d\right)/c$. Clearly, a dependence on the sign of $k_z$ (or equivalently $\beta$) arises in the presence of a magnetic field (i.e., when $\epsilon_{xz}\neq0$). Incidentally, we can separate the surface mode [Fig.\ \ref{Fig2}(b), red curves], which in the large $d$ limit [i.e., when the modes does not reach the aluminum substrate, as shown in panel (1) of Fig.\ \ref{Fig2}(b)] reduces to \cite{DE13}
\begin{align}
\sqrt{\beta^{2}-1}+\frac{\sqrt{\beta^{2}-\epsilon_{1}}}{\epsilon_{1}}=\frac{-\ii\beta\epsilon_{xz}}{\epsilon_{1}\epsilon_{zz}}. \nonumber
\end{align}
The electron line [$\omega=k_{z}v$, vertical black dotted line in Fig.\ \ref{Fig2}(b)] crosses the dispersion curves at multiple frequencies, which perfectly match the spectral positions of the resonance peaks shown in Fig.\ \ref{Fig1}(b), confirming again their physical origin. Naturally, by changing the electron velocity $v$, the crossings of the dispersion curves take place at a different set of frequencies, which produce the corresponding resonance peaks in the EELS signal (see supplementary Fig.\ \ref{FigS2}). 

\begin{figure}
\centering{\includegraphics[width=0.5\textwidth]{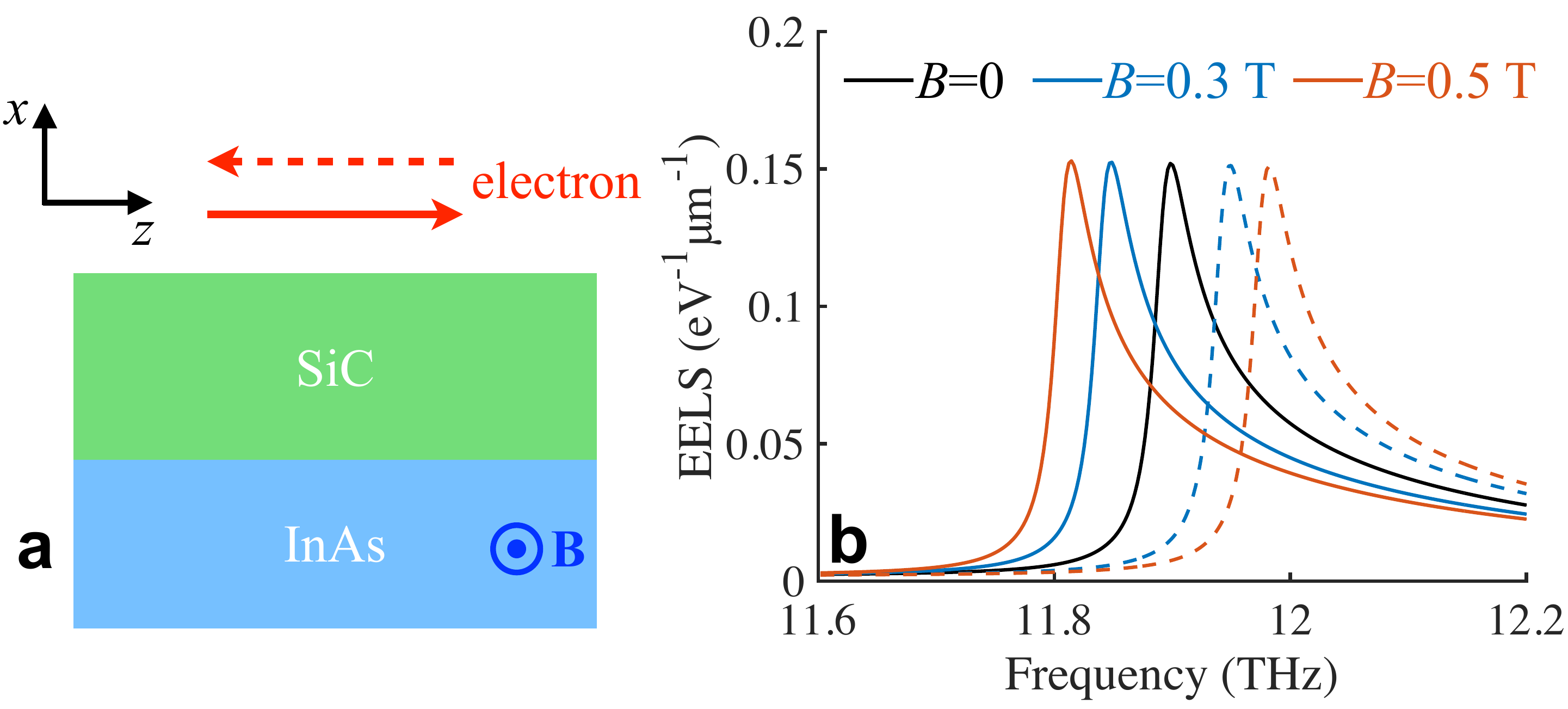}}
\caption{{\bf Probing the nonreciprocal response of a buried substrate.} {\bf (a)} We consider an InAs planar substrate coated with a 4.65-{\textmu}m-thick layer of SiC. Electrons are moving parallel to the film at a distance of 1\,{\textmu}m from the upper surface. An external magnetic field $\Bb$ is applied with the orientation shown in the figure. {\bf (b)} Simulated EELS probability spectra for electrons moving with velocity $v=0.5\,c$ to the right (solid curves) or to the left (broken curves) for different strengths of the applied magnetic field (see color-matched labels).}
\label{Fig3}
\end{figure}

In order to reduce the magnitude of the external magnetic field needed to obtain a strong nonreciprocity, we study waveguide modes with higher quality factors. Following previous work \cite{ZSW19}, we consider a 4.65\,\textmu m waveguide of silicon carbide (SiC, low-loss material, permittivity from Ref.\ \citenum{ZSW19}) supported on InAs, as shown in Fig.\ \ref{Fig3}(a). Nonreciprocity is produced by a transverse magnetic field acting on the InAs substrate, as a significant fraction of the waveguide mode fields penetrate inside this material (see supplementary Fig.\ \ref{FigS1}). Specifically, we explore a frequency range around the $\rm TM_1$ mode of the SiC waveguide (see near-field distribution in Fig.\ \ref{FigS1}). Similar to the results presented in Fig.\ \ref{Fig1}(b), we observe in Fig.\ \ref{Fig3}(b) a sizable resonance splitting for electrons moving along opposite directions, but this happens for a smaller magnitude of the magnetic field down to 0.3\,T. To make the comparison more quantitative, we define an asymmetry parameter $\zeta=|{\rm EELS}(\omega_{\rm res},k_z)-{\rm EELS}(\omega_{\rm res},-k_z)|/{\rm EELS}(\omega_{\rm res},k_z)/B$, where the mode resonance frequency $\omega_{\rm res}=\omega_{\rm TM_1}(k_z)$ is determined by the longitudinal wave vector transfer $k_z=\omega_{\rm res}/v$. For the $\rm TM_1$ resonance in Fig.\ \ref{Fig1}(b), we have $\zeta=0.961/{\rm T}$, whereas $\zeta=3.204/$T at $B=0.3\,$T in Fig.\ \ref{Fig3}(b). Incidentally, we find $\omega_{\rm TM_1}(k_z)<\omega_{\rm TM_1}(-k_z)$ in Fig.\ \ref{Fig3}(b), whereas the opposite is true in Fig.\ \ref{Fig1}(b), a result that can be related to the fact that the nonreciprocal material is outside or inside the waveguide, respectively.

\begin{figure*}
\centering{\includegraphics[width=0.85\textwidth]{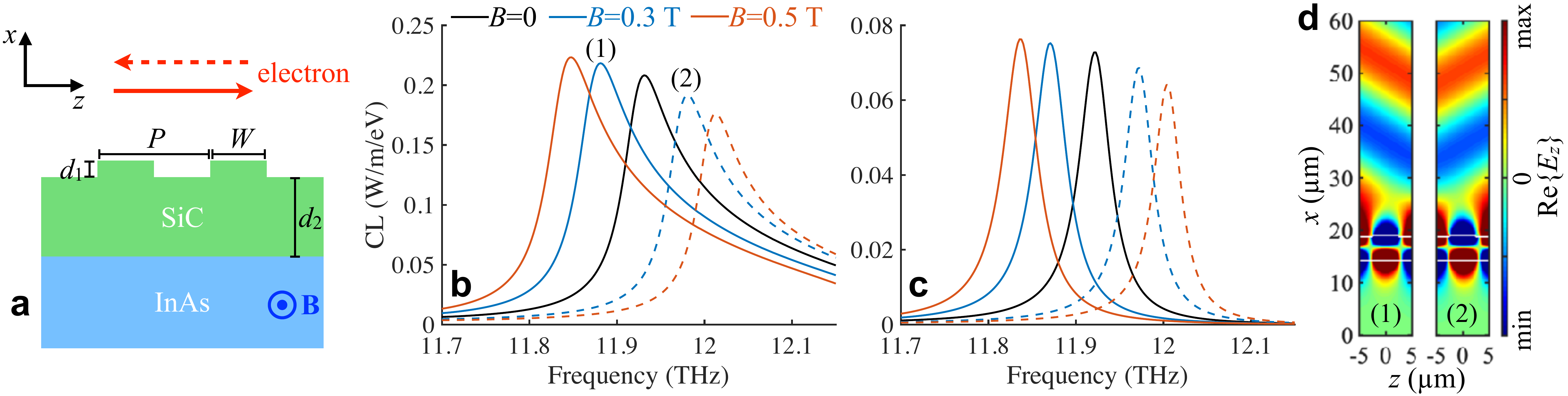}}
\caption{{\bf Nonreciprocal CL emission.} {\bf (a)} We consider a structure similar to that of Fig.\ \ref{Fig3}(a), with a grating carved in the upper surface to produce mode leakage to emitted light. The geometrical parameters indicated in the figure are $P=10\,${\textmu}m, $W=5\,${\textmu}m, $d_{1}=0.3\,${\textmu}m, and $d_{2}=4.5\,${\textmu}m. {\bf (b,c)} Simulated CL probability spectra for electrons moving with velocity $v=0.5\,c$ as indicated in (a) either to the right (solid curves) or to the left (broken curves) for different strengths of the applied magnetic field (see color-matched labels) and 1\,{\textmu}m distance to the top grating surface. The emission is integrated over the full upper hemisphere in (b) and up to angles $<5^\circ$ relative the $y=0$ plane in (c). {\bf (d)} Electric near-field amplitude distributions corresponding to the modes (1) and (2) indicated in (b) for $B=0$. White lines delineate one period of the SiC grating.}
\label{Fig4}
\end{figure*}

By patterning a grating on the upper surface of the SiC layer [Fig.\ \ref{Fig4}(a)], waveguide modes are out-coupled to CL emitted light. We calculate the CL signal by integrating the Ponyting vector of the emitted light in the far field, from which we observe again a splitting of the waveguide resonance in the CL spectra for electrons moving along opposite directions perpendicular to an externally applied DC magnetic field [Fig.\ \ref{Fig4}(b)]. The corresponding spectral positions of the resonance peaks match very well with those in Fig.\ \ref{Fig3}(b), thus revealing their common physical origin, which is the excitation of waveguide modes whose electric field distributions inside the SiC layer [see Fig.\ \ref{Fig4}(c) for the two peaks labelled in Fig.\ \ref{Fig4}(b)] are very similar to those shown in supplementary Fig.\ \ref{FigS1} for the planar SiC structure: the grating does not distort these modes substantially, other than to produce out-coupling to emitted light. For electrons traveling along those two opposite directions, light is emitted along different preferential directions derived from umklapp scattering of the guided modes by the grating, so that the corresponding emission angles $\theta_{+}$ and $\theta_{-}$ (with respect to the positive {\textit x} axis) satisfy the condition $\theta_{+}+\theta_{-}=\pi$ for $B=0$, but not in the nonreciprocal structure driven by a finite magnetic field.

\section*{CONCLUSIONS}

In conclusion, we have demonstrated that free electrons can probe the nonreciprocal response of magnetically biased photonic structures based on gyrotropic materials. We have shown that the observed differences in EELS and CL spectra are apparent for electrons propagating along two opposite directions. We attribute the splitting of multiple waveguide resonance peaks in the electron-generated spectra to the asymmetric dispersions with respect to the mode propagation direction. Such asymmetry can be tuned by changing the magnitude of the applied magnetic field, which introduces modulation of the observed spectral features. In particular, we have shown that high-quality-factor waveguides can produce strong nonreciprocal electron-matter interactions with relatively moderate applied magnetic fields. By varying the strength of the magnetic field, we have demonstrated tuning of cathodoluminescence light emission from nonreciprocal gratings, thus providing an active way of control for e-beam-based light sources. We note that similar nonreciprocal interaction should also be observable using other nonmagnetic systems, such as time-varying photonic resonators \cite{SA14} or Weyl semimetals \cite{HS16}, which deserve further exploration. Our findings demonstrate the ability of e-beams to sample direction-dependent nonreciprocal response and the possibility of tuning e-beam-based light sources through externally applied magnetic fields.






\begin{acknowledgments} 
This work has been supported in part by the European Research Council (Advanced Grant 789104-eNANO), the European Commission (Horizon 2020 Grants 101017720 FET-Proactive EBEAM and 964591-SMART-electron), the Spanish MINECO (Severo Ochoa CEX2019-000910-S), the Catalan CERCA Program, and Fundaci\'{o}s Cellex and Mir-Puig. 
\end{acknowledgments} 

\bibliographystyle{apsrev} 
\bibliography{../../../bibtex/refsL.bib} 

\clearpage 
\pagebreak \onecolumngrid \section*{SUPPLEMENTARY FIGURES} 

\begin{figure*}[h]
\begin{centering}
\includegraphics[width=0.7\textwidth]{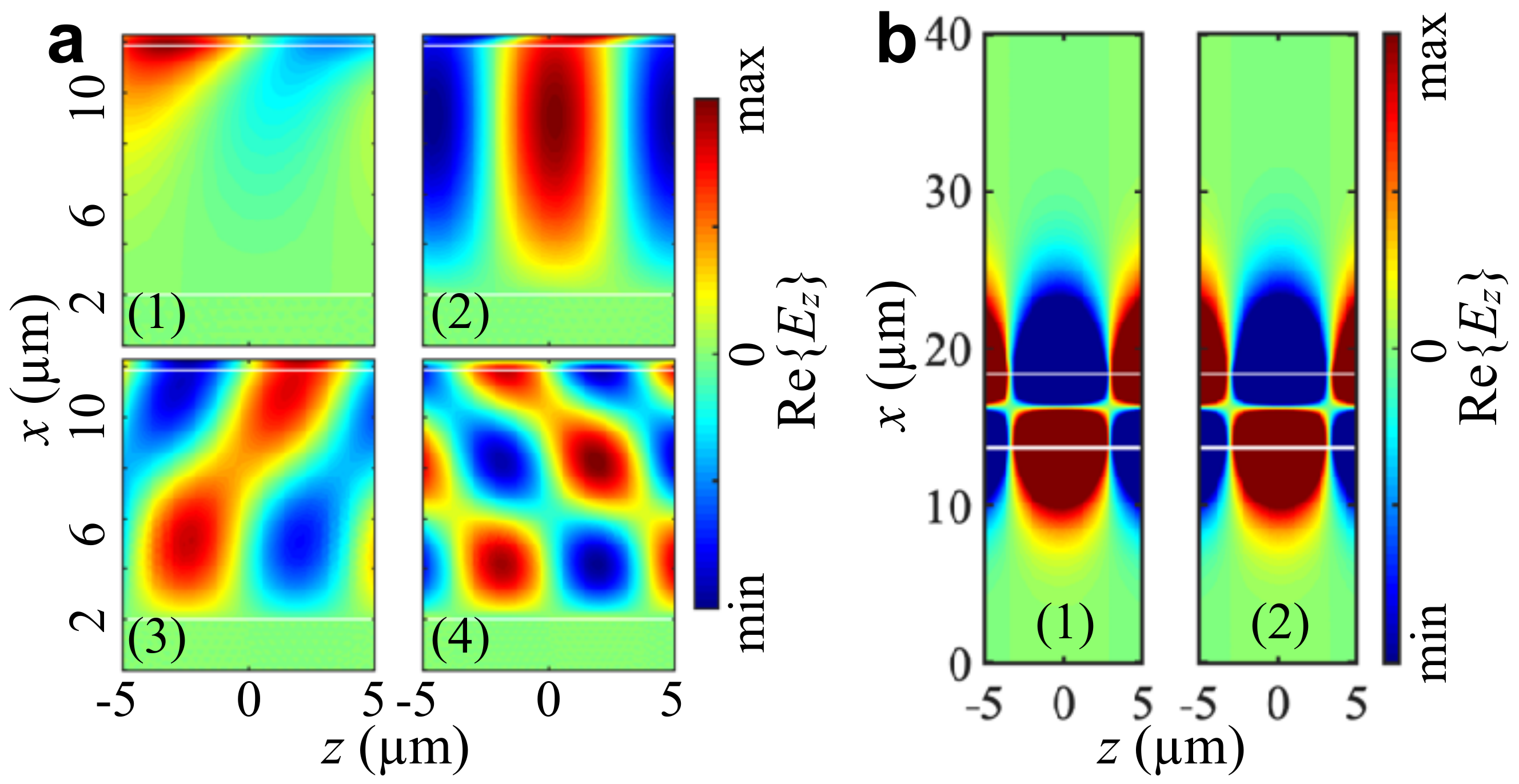}
\par\end{centering}
\caption{(a) Difference between the electric field distributions between the left- and right-propagating split resonance peaks shown in Fig.\,1(b) (see corresponding labels there) with a applied DC magnetic field of strength $B=1\,$T. (b) Same as Fig.\ 4(c), but in the presence of a magnetic field of strength $B=0.3\,$T.}
\label{FigS1}
\end{figure*}

\begin{figure*}
\begin{centering}
\includegraphics[width=0.8\textwidth]{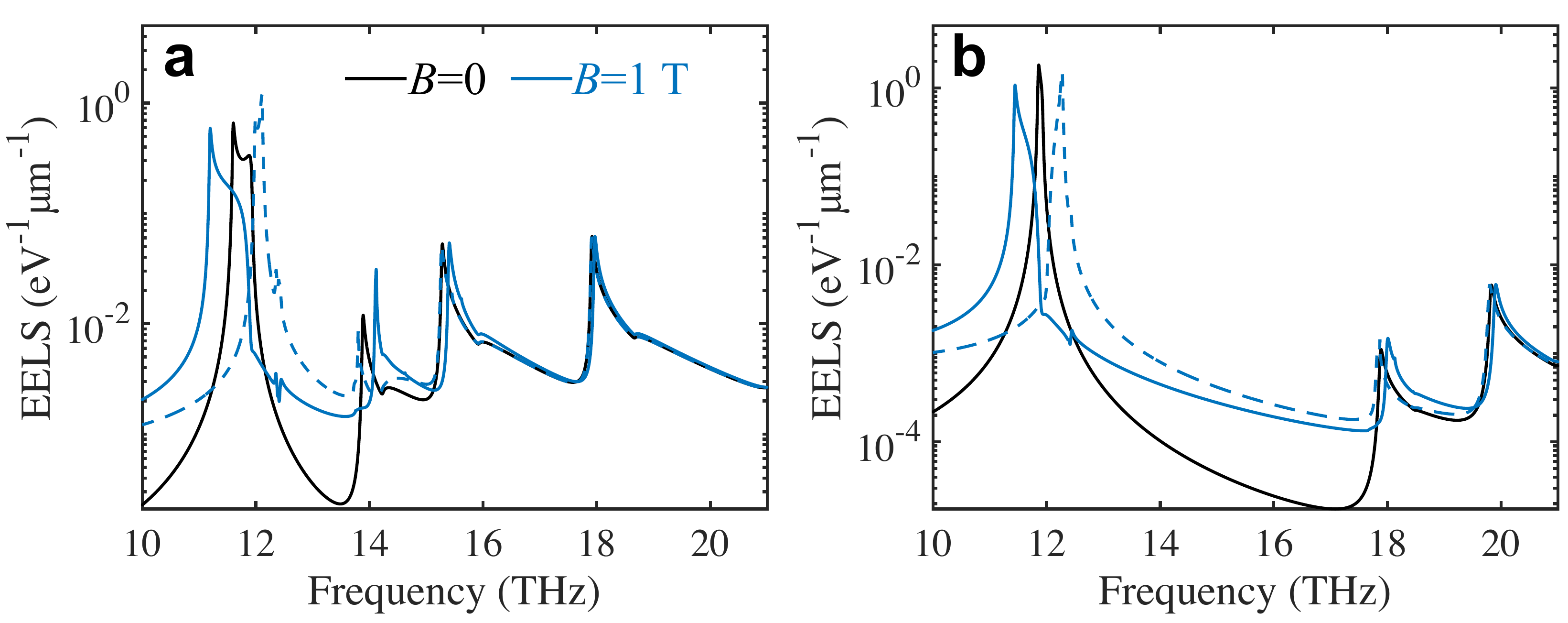}
\par\end{centering}
\caption{Calculated EELS probability for the same configuration as depicted in Fig.\,1(a), but for different electron velocities: (a) $v=0.67\,c$ and (b) $v=0.4\,c$.}
\label{FigS2}
\end{figure*}

\end{document}